\documentclass{article}
\usepackage{waspaa23,amsmath,graphicx,url,times}
\usepackage{color}

\usepackage{ifthen}
\usepackage{multicol,multirow}
\usepackage{booktabs}

\usepackage{xcolor}
\usepackage{kotex}

\title{All-In-One Metrical And Functional Structure Analysis\\With Neighborhood Attentions on Demixed Audio}

\name{Taejun Kim
      and Juhan Nam\thanks{This research was supported by Culture, Sports and Tourism R\&D Program through the Korea Creative Content Agency grant funded by the Ministry of Culture, Sports and Tourism in 2023 (Project Name: Development of high-speed music search technology using deep learning, Project Number: CR202104004)}}
\address{KAIST, Graduate School of Culture Technology, Daejeon, Republic of Korea}

\begin{document}

\ninept
\maketitle

\begin{sloppy}

\begin{abstract}
Music is characterized by complex hierarchical structures. Developing a comprehensive model to capture these structures has been a significant challenge in the field of Music Information Retrieval (MIR). Prior research has mainly focused on addressing individual tasks for specific hierarchical levels, rather than providing a unified approach. In this paper, we introduce a versatile, all-in-one model that jointly performs beat and downbeat tracking as well as functional structure segmentation and labeling. The model leverages source-separated spectrograms as inputs and employs dilated neighborhood attentions to capture temporal long-term dependencies, along with non-dilated attentions for local instrumental dependencies. Consequently, the proposed model achieves state-of-the-art performance in all four tasks on the Harmonix Set while maintaining a relatively lower number of parameters compared to recent state-of-the-art models. Furthermore, our ablation study demonstrates that the concurrent learning of beats, downbeats, and segments can lead to enhanced performance, with each task mutually benefiting from the others.
\end{abstract}


\section{Introduction}
\label{sec:intro}
Music has a hierarchical organization characterized by distinct levels of structural units. The foundational level comprises metrical elements, including beats, bars, and segments, which form the basic rhythmic structure. Ascending the hierarchy, these metrical components are assembled into functional units, such as verses and choruses, that collectively shape the overall architecture of the piece. Despite inherent interdependence of the hierarchical levels, research in the field of MIR has primarily been conducted as isolated tasks such as beat/downbeat tracking~\cite{bock2016joint,bock2019multi,bock2020deconstruct,chen2022toward,zhao2022beat}, segmentation~\cite{mccallum2019unsupervised,salamon2021deep}, and functional structure labeling~\cite{nieto2020audio,wang2021supervised,wang2022catch}, missing the potential benefits of interdependence gained from all the metrical and functional structure information. 
However, joint learning of the hierarchical information levels in a unified model presents considerable challenges due to the substantial length and high dimensionality of individual songs represented as audio data. Furthermore, the songs contain a wide variety of acoustic and musical variations within the underlying metrical and functional structure layers. In this paper, we attempt to predict beat, downbeat, segmentation, and functional structure labels all at once with a single model and show their synergy in the multi-task learning.   

The core challenge in the attempt is designing an efficient model that can learn the information with a large time-granularity over long-range audio frame sequences. In the beat/downbeat tracking task, the model has been designed to have a large receptive field to cover a sufficient number of beats and downbeats. A representative model is Temporal Convolutional Networks (TCN), a family of convolutional neural networks with dilation operations which has an exponentially increasing size of receptive fields as the layer goes up~\cite{bock2016joint,bock2019multi,bock2020deconstruct,chen2022toward}. Recently, researchers have improved the performance further using variants of the transformer architecture. For example, SpecTNT-TCN used the time-frequency transformer (SpecTNT) for efficient long-term representation learning and integrated it with the TCN module for performance gain~\cite{hung2022modeling}. 
Beat Transformer employed the dilation operations in the self-attention layers along with demixed input, achieving state of the art results across five datasets~\cite{zhao2022beat}. 

Unlike beats and downbeats, segmentation boundaries and  temporal change of functional structure labels are much sparser. Thus, the tasks have primarily been tackled as segmentation problems based on the self-similarity of local audio features within a song~\cite{nieto2020audio}. One group of previous works explored better audio features or embeddings using temporal affinity~\cite{mccallum2019unsupervised}, semantic labels~\cite{salamon2021deep}, or structure labels~\cite{wang2021supervised}. The other group focused on segmentation algorithms that leverage homogeneity, repetition, and novelty principles in the segment level~\cite{Foote2002,mcfee2014learning,Maezawa2019}. However, recent models based on convolutional neural networks or transformer predicted the ``boundaryness'' or  ``chorusness'' of an excerpt directly from the audio and achieved a new state-of-the-art~\cite{wang2021chorus,wang2022catch}.

Following recent advances in the aforementioned tasks, our proposed model builds upon the transformer architecture. Specifically, we incorporate dilated self-attention layers and demixed input from Beat Transformer. However, we introduce three major modifications. First, we employ ``neighborhood attention'' which effectively creates attention windows enclosing nearest possible neighbors without requiring zero-padding~\cite{hassani2023neighborhood}. This facilitates widening the receptive field of the model without unnecessary computation. Second, we set the model to predict not only beat and downbeat but also segmentation boundary and functional structure labels directly from audio input. Through a comprehensive ablation study, we investigate performance interaction in the all-in-one learning. Lastly, we significantly streamline the model size following the configuration of the TCN model. We evaluated the proposed model on the Harmonix Set which includes all metrical and functional structure labels~\cite{nieto2019harmonix}. We show that our proposed model outperforms recent state-of-the-art models in all four tasks while maintaining a relatively small number of parameters (about 300K). The code and pre-trained models are accessible via the provided link~\footnote{\url{https://github.com/mir-aidj/all-in-one}}.

\begin{figure*}[!t]
\label{fig:arch}
  \centering
  \centerline{\includegraphics[width=.91\linewidth]{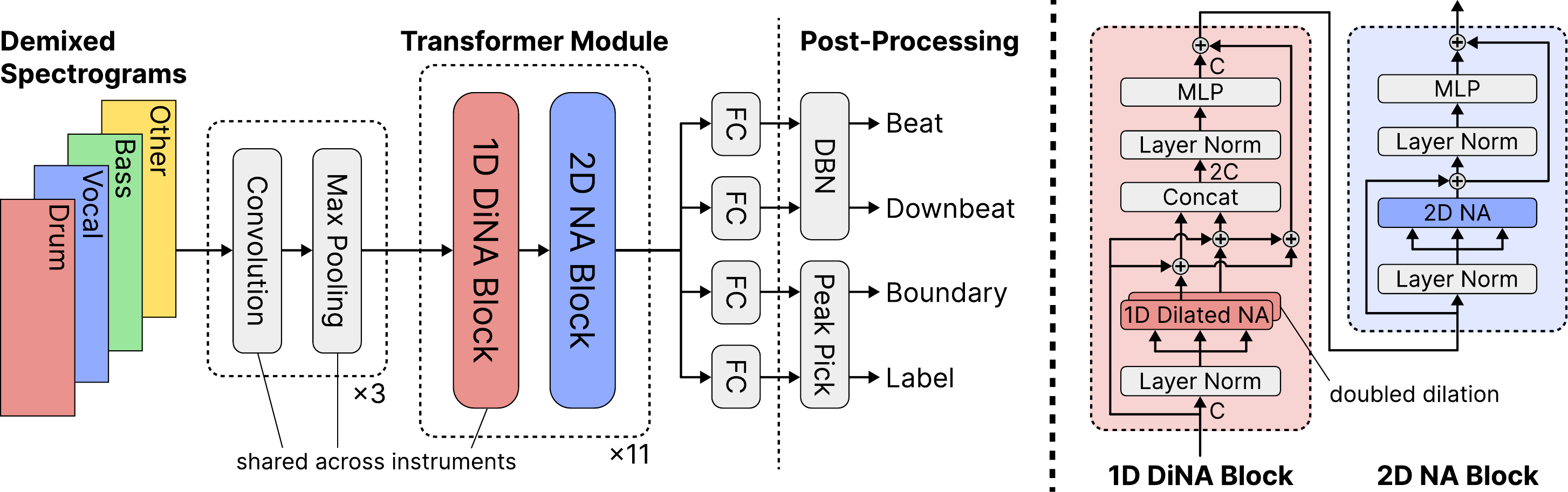}}
  \caption{(Left) An illustration of the proposed model architecture. (Right) A detailed representation of the transformer module, showcasing both the 1-dimensional (1D) Dilated Neighborhood Attention (DiNA) and the 2-dimensional (2D) Neighborhood Attention (NA) blocks. C denotes the embedding dimension.}
\end{figure*}

\section{Method}
\label{sec:method}

\subsection{Model Architecture}
An overview of the proposed model is illustrated in the left side of the \figurename~\ref{fig:arch}. The model utilizes demixed sources as inputs and convolutional layers and max-pooling as a front end processing following Beat Transformer~\cite{zhao2022beat}. However, the transformer modules comprise two distinct blocks based on with neighborhood attentions: 1) 1D Dilated Neighborhood Attention (DiNA) block which models long-term temporal dependencies using dilations, and 2) 2D Neighborhood Attention (NA) block which models inter-instrument dependencies while preserving locality by focusing on local neighbors. The concept of stacking alternating dilated and non-dilated blocks originates from the original DiNA design~\cite{hassani2022dilated}.   


The 1D DiNA block includes two DiNA modules inspired by the TCN model~\cite{bock2020deconstruct}. The proposed transformer module has the second DiNA module with a doubled dilation, aiming for the model to learn musical properties at various levels that are integer multiples of each other. The outputs of the two DiNA modules are first added to the skip connection, then concatenated, and fed into the next layer. The multilayer perceptron (MLP) consists of two fully connected layers, which initially increase the embedding dimension to $8C$ and subsequently reduce it back to its original size of $C$ to keep the embedding size consistent. The dilations grow up to $2^{10}$ and $2^{11}$, yielding receptive field sizes of approximately 41 and 82 seconds for the first and second DiNA modules, respectively. The size of the embedding dimension $C$ remains fixed at 24 throughout all transformer blocks. The 2D NA block is identical to the original NA~\cite{hassani2023neighborhood}.

\begin{figure}[t]
  \centering
  \centerline{\includegraphics[width=\columnwidth]{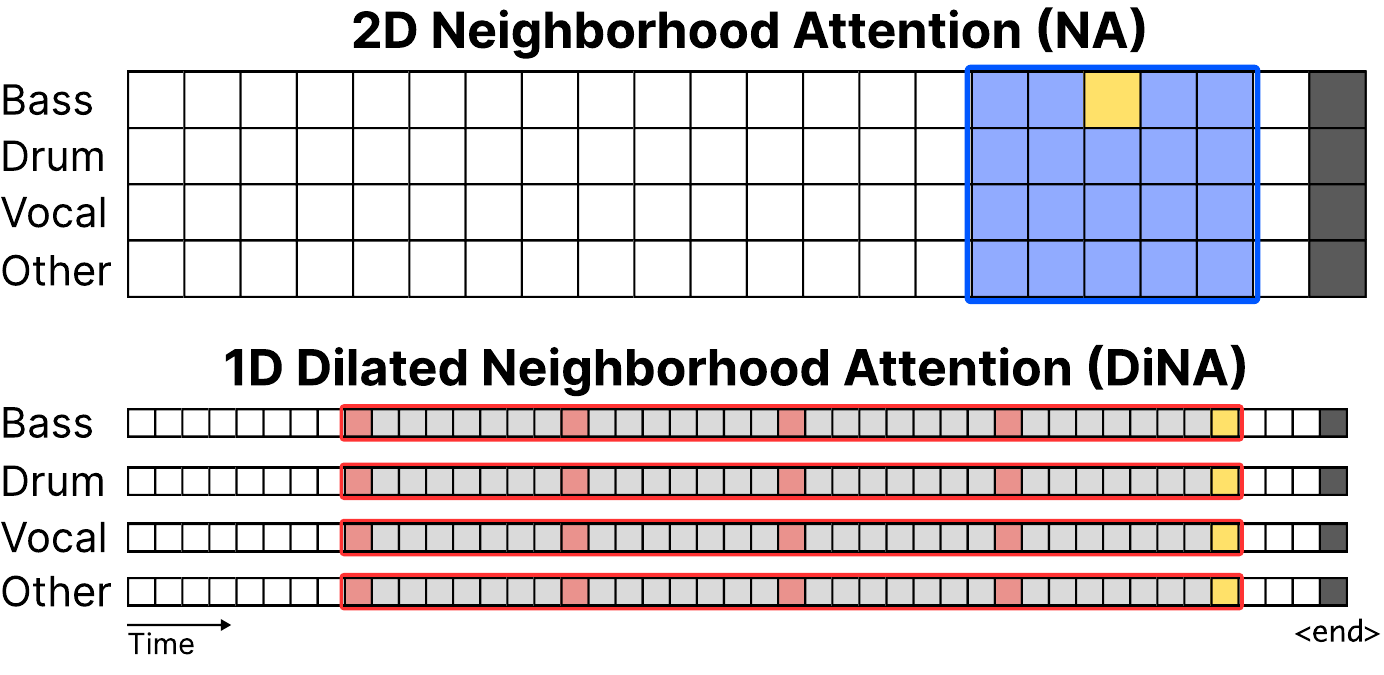}}
  \caption{An illustration of the attention windows (depicted by red and blue lines) in Neighborhood Attentions~\cite{hassani2022dilated,hassani2023neighborhood} at the end of a song. Unlike conventional sliding window attention or convolution mechanisms, the windows are not centered around the attending (yellow; or query) values. Instead, they enclose the nearest possible neighbors (red and blue boxes), effectively eliminating the need for zero padding. The light grey boxes represent the dilations.}
  \label{fig:attn}
\end{figure}

\begin{figure}[!t]
  \centering
  \hspace{-15pt}
  \includegraphics[width=.98\columnwidth]{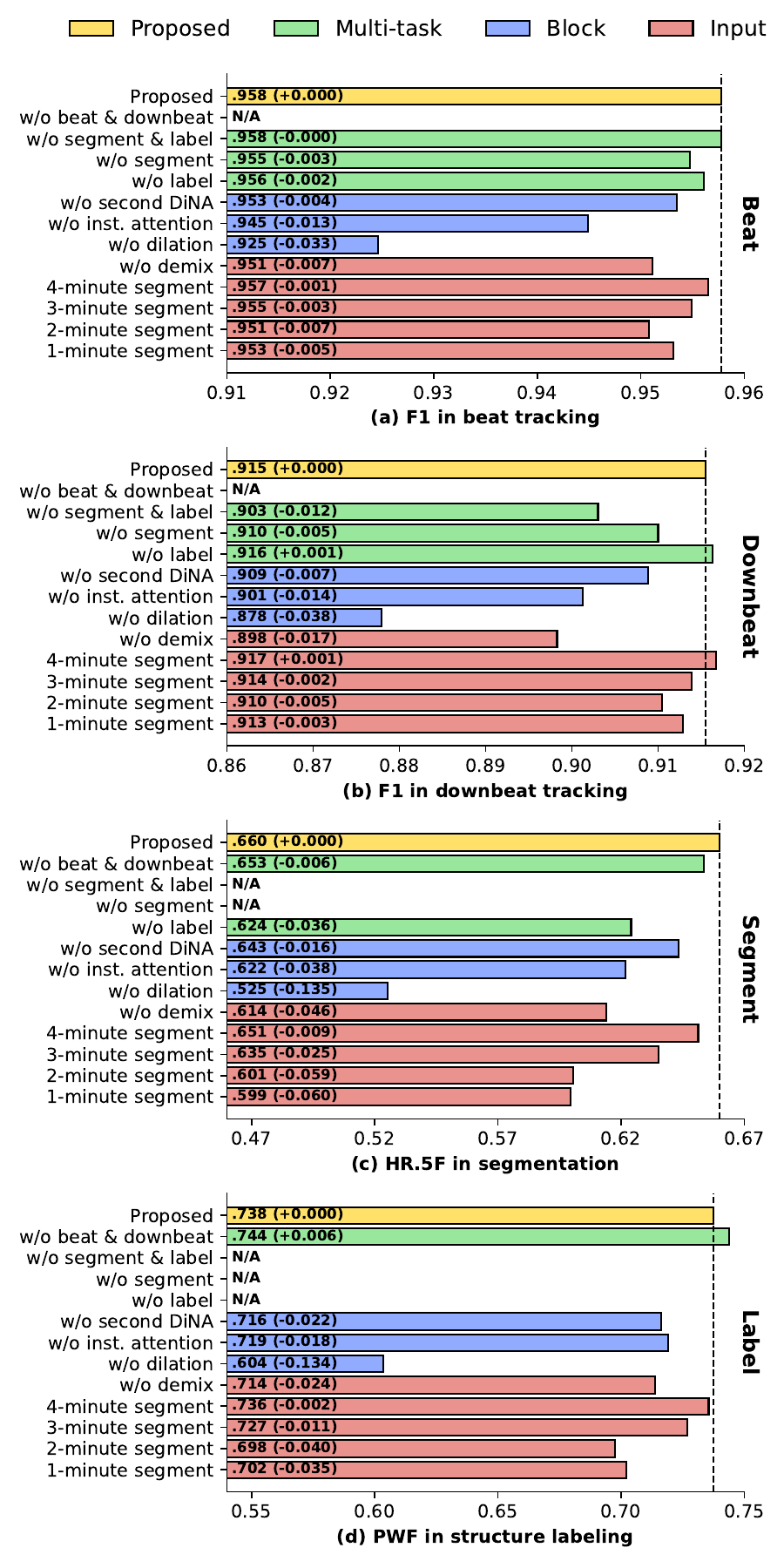}
  \caption{Ablation study performance results. Performance differences relative to the proposed model (dotted line) are indicated in parentheses.}
  \label{fig:abl}
  \vspace{-4mm}
\end{figure}

\subsection{Details of Neighborhood Attentions}
\figurename~\ref{fig:attn} illustrates the neighborhood attention mechanism at the end of a song.
The bottom part of the figure demonstrates how the DiNA effectively and efficiently computes attention at the end of a song without requiring any zero padding. 
In the worst cases, with a large receptive field such as 82 seconds, conventional mechanisms would require 41 seconds of zero padding, which adds unnecessary computational complexity.
The top part of the figure shows that the NA effectively creates the window only enclosing available instruments and time frames surrounding it\footnote{In practice, we use a kernel size of $5\times5$ with a zero padding on the instrumental dimension since NATTEN does not support non-square kernels.}.
These details make the proposed model different from the Beat Transformer, which has fixed sliding window and single frame instrumental attention.


\subsection{Model Configuration and Post-processing}
The transformer architecture generally requires a large number of parameters. Inspired by the effectiveness of lightweight TCN models~\cite{bai2018tcn}, we streamline the proposed transformer model. Specifically, we followed the overall configuration and pipeline from the TCN models for beat, downbeat, and tempo estimation~\cite{bock2019multi,bock2020deconstruct}. We adopt the same input spectrogram configurations and initial feature extractor setups, which consist of three convolutional and max pooling layers. Our proposed model also has a stack of 11 sequence modeling blocks (transformer modules in this work) and utilizes a dynamic Bayesian network (DBN)~\cite{krebs2015efficient} for post-processing beats and downbeats. However, since the TCN model is designed solely for beat and downbeat tracking, we apply the peak-picking method from two other previous works~\cite{ullrich2014boundary,wang2022catch} for post-processing segment boundaries and functional labels. This method involves normalizing the probabilities of segment boundaries using sliding window averages and selecting the highest probability. Contrary to the previous works, we do not apply thresholding after normalization and opt for a window size of 24 seconds, as opposed to their 18-second window. In the sequence modeling block design, further adaptations from the TCN model are made: a kernel size of 5, a second kernel featuring doubled dilation, and an exponentially increasing dilation rate at $2^l$, where $l$ represents the block number.

\section{Experiments}

\subsection{Experimental Setup}

We primarily used the Harmonix Set~\cite{nieto2019harmonix} for the experiment. We conducted data cleaning and functional label merging following the previous work~\cite{wang2022catch}, as different versions of audio and annotations exist. The labels represent segment functions such as `verse' and `chorus'. Performance evaluation is carried out under 8-fold cross validation, following the convention of beat and downbeat tracking~\cite{matthewdavies2019temporal,bock2020deconstruct,hung2022modeling,zhao2022beat}. Among the 8 folds, 6 are designated for training, 1 for validation, and 1 for test. 
Data augmentation and additional datasets are not utilized in this study.

\subsection{Evaluation Metrics}
We assess performance using conventional metrics for each task. For beat and downbeat tracking, F-measure (F1) with a tolerance window of 70 ms, CMLt, and AMLt are utilized~\cite{davies2009evaluation}. For segmentation, the F-measure of hit rate at 0.5 seconds (HR.5F) is employed, while the F-measure of pairwise frame-level clustering (PWF) and F-measure of normalized entropy score (Sf) are used for the evaluation of segment labeling~\cite{nieto2020audio}.

\subsection{Implementation Details}

We reproduced the TCN model on our own according to the original training strategies and hyperparameters~\cite{bock2020deconstruct}. However, for the larger variation (TCN-Large), we perform a grid search to determine the optimal regularization hyperparameters.
TCN-Large is a variant designed to offer a fair comparison with the proposed model, as it has a similar number of parameters (301 K) by increasing the channel dimensionality.
We use PyTorch 2.0 for implementation. Hybrid Transformer Demucs~\cite{rouard2022hybrid} handles source separation, while NATTEN\footnote{\url{https://github.com/SHI-Labs/NATTEN}} implements NA and DiNA.  Madmom~\cite{madmom} is utilized for spectrogram extraction and DBN implementation. The batch size is set to 1, and spectrograms larger than 5 minutes are randomly chunked into 5 minutes due to the GPU memory limit. Optimization is performed using RAdam~\cite{Liu2020On} with a learning rate of 0.005 and Stochastic Weight Averaging (SWA)~\cite{izmailov2018averaging} with a learning rate of 0.15. When the validation loss plateaus, the learning rate is decayed by a factor of 0.3. A weight decay of 0.00025 is applied. Early stopping is triggered when the validation loss fails to decrease for 30 epochs. Dropouts with rates of 0.2, 0.2, 0.2, and 0.1 are applied to convolution, MLP, attention probabilities, and skip connections. Exponential Linear Unit is utilized for convolutions and Gaussian Error Linear Unit for transformers. On average, early stopping is activated after 5 hours of training on a single fold with an RTX 2080 Ti 11 GB.

\begin{table*}[!t]
\centering
\resizebox{.85\textwidth}{!}{
\begin{tabular}{lrccccccccccccc}
\toprule
\multirow{2}{*}[-.5ex]{Model} & \multirow{2}{*}[-.5ex]{\shortstack{\# of \\ Params}} & \multicolumn{3}{c}{Beat} && \multicolumn{3}{c}{Downbeat} && Segment && \multicolumn{2}{c}{Label} \\
\cmidrule{3-5}
\cmidrule{7-9}
\cmidrule{11-11}
\cmidrule{13-14}
& & F1 & CMLt & AMLt && F1 & CMLt & AMLt && HR.5F && PWF & Sf \\
\toprule

SpecTNT-TCN~\cite{hung2022modeling}$^\ast$         & 4.7 M & .953 & \textbf{.939} & .959 && .908 & .872 & .928 &&  --  &&  --  &  --  \\
Beat Transformer~\cite{zhao2022beat}$^\ast$        & 9.3 M & .954 & .905 & .957 && .898 & .863 & .919 &&  --  &&  --  &  --  \\ 
\midrule
DSF+Scluster~\cite{wang2021supervised}             &  N/A  &  --  &  --  &  --  &&  --  &  --  &  --  && .497 && .684 & .743 \\
SpecTNT~\cite{wang2022catch}$^\ast$                &  N/A  &  --  &  --  &  --  &&  --  &  --  &  --  && .558 && .712 & .724 \\
\midrule
TCN w/o demix~\cite{bock2020deconstruct}$^\dagger$ &  74 K & .954 & .900 & .961 && .886 & .842 & .920 && .594 && .687 & .694 \\
TCN$^\dagger$                                      &  93 K & .946 & .898 & .950 && .894 & .850 & .919 && .619 && .715 & .738 \\
TCN-Large$^\dagger$                                & 301 K & .953 & .906 & .960 && .901 & .853 & .924 && .626 && .717 & .746 \\

\midrule
All-In-One-Small~(Ours)                            &  46 K & .943 & .891 & .952 && .901 & .854 & .929 && .616 && .713 & .745 \\
All-In-One~(Ours)                                  & 300 K & \textbf{.958} & .913 & \textbf{.964} && \textbf{.915} & \textbf{.873} & \textbf{.932} && \textbf{.660} && \textbf{.738} & \textbf{.769} \\
\bottomrule
\end{tabular}
}
\caption{Comparison of performance metrics between previous works and the proposed models on the Harmonix Set. $^\dagger$denotes previous works reproduced by us and their number of parameters are also calculated by us. $^\ast$indicates the use of data augmentation and extra datasets.}
\label{tab:results}
\end{table*}

\section{Results and Discussion}

\subsection{Ablation Study}
To investigate the contributions of various components to the overall performance gain, we conduct an ablation study by discarding a component or changing training setups, as shown in \figurename~\ref{fig:abl}. The ablation study is grouped into modifications of multi-task, block, and input settings. They are colored in green, blue, and red, respectively. All performance results are averages of 8-fold cross-validation results, and a single metric for each task is used as a representative metric: F1, F1, HR.5F, and PWF, for beat, downbeat, segment, and label, respectively.

\textit{Multi-task settings} are cases where specific losses are discarded. For example, ``w/o beat \& downbeat'' does not including beats and downbeat tracking tasks but only focuses on segmentation and structure labeling, resulting in the absence of performance metrics for the beat and downbeat tracking (shown as ``N/A''). The downbeat tracking performance significantly decreases without the segmentation tasks and structure labeling as shown in \figurename~\ref{fig:abl} (b) but we found that this is due to overfitting by investigating the training and validation loss curves. 
Nonetheless, it seems that the three tasks (beat/downbeat tracking and segmentation) benefit from the joint learning. Beat and downbeat tracking performances decrease without the segmentation task, and vice versa. However, they are not influenced by the joint learning with structure labeling. Furthermore, when considering the downbeat tracking and structure labeling performances, their performances become higher without each other. This may be because the nature of structure is more akin to long-term timbre classification rather than instant event detection such as the three other tasks. Nevertheless, segmentation performance drops without learning of the structure labeling because it can provide strong cues to find segment boundaries when the labels change.

\textit{Block settings} refer to training models with the omission of one component in the transformer module at a time. The ``w/o second DiNA" setting lacks the second 1D Dilated Neighborhood Attention (NA) with doubled dilation (as depicted on the right side of \figurename~\ref{fig:arch}). In the ``w/o inst. attention" setting, the model uses 1D NA instead of 2D NA, which means the models do not have any instrument-wise attention. Models labeled as ``w/o dilation" do not have any dilations. The results from these experiments demonstrate that performance decreases when any of the components in the block are missing, highlighting the importance of each component in the transformer module for achieving optimal performance.

\textit{Input settings} refer to different settings of input channel and length. ``w/o demix'' indicates models without demix inputs, which drastically decreases performances in all four tasks. Lastly, we provide performance metrics for models trained with shorter segment lengths and higher batch sizes. For example, ``1-minute segment" indicates models trained with randomly chunked 1-minute spectrograms. To take advantage of shorter chunks, we set the batch sizes to 5, 3, 1, and 1 for 1, 2, 3, and 4-minute segments, respectively, which are the maximum numbers that can be loaded onto the GPU. While it is known that a large batch size leads to better generalization, it is impossible to achieve if the sequence length is too long due to the limited memory of the GPU. However, the results show that training with longer segments and a batch size of 1 yields better generalizations, especially in segmentation and structure labeling. This is probably because the long-term nature of segments and structure labels requires a large context to predict.

\subsection{Comparison with Previous Work}
\tablename~\ref{tab:results} provides a summary of the performance of current state-of-the-art models in beat/downbeat tracking, segmentation, and structure labeling. Meanwhile, the bottom two sections display the performance of the TCNs and the proposed models. The TCNs are evaluated with and without demixed inputs. The proposed model (All-In-One) outperforms the TCNs models. Additionally, we can see that demixing is effective for the TCNs as well except for beat tracking. Notably, the proposed model achieves state-of-the-art performances in all four tasks while maintaining a relatively small number of parameters compared to other models. We also report the performances of a smaller version of the proposed model, which only has 46 K parameters. This smaller model consists of nine stacks of the transformer module, a kernel size of three, an embedding size of 16, and exponentially growing dilations with a factor of three. Remarkably, even with the small number of parameters, the proposed model already achieves state-of-the-art performances in the segmentation and structure labeling tasks.

\section{Conclusions}
We introduced a novel approach that learns multiple levels of hierarchical music structure, including beats, downbeats, segment boundaries, and functional labels. To construct the model, we employed demixed sources as inputs and adopted neighborhood attentions for effective modeling of temporal and instrumental dependencies. As a result, the proposed model achieved state-of-the-art performances in all four tasks. Furthermore, our ablation study reveals the potential benefits of joint training for beats, downbeats, and segments, while structure labels may not derive the same advantage. We hypothesize that the reason for this observation is that that structure labeling focuses on long-term timbre texture whereas the other three tasks involve short-term event detection. 


\bibliographystyle{IEEEtran}
\bibliography{refs23}
%
%
%
%
%
%
%
%
%

\end{sloppy}
\end{document}